\documentclass{article}
\usepackage{spconf,amsmath,graphicx}
\usepackage[textfont=it,tableposition=top]{caption}
\usepackage{hyperref}
\hypersetup{colorlinks=true}
\usepackage{multirow}
\usepackage[hang,flushmargin]{footmisc}

\def \cleanCgm          {\mathbf X}
\def \EstCleanCgm       {\widehat{\cleanCgm}}
\def \noiseCgm          {\mathbf N}

\def \mixCgm            {{\mathbf Y}}   %

\def \IRM               {M}           %
\def \ERM               {\widehat{M}}   %
\def \TI                {t}             %
\def \FI                {c}

\renewcommand{\paragraph}[1]{\textit{#1}\quad}

\title{A Conformer-based ASR Frontend for Joint Acoustic Echo Cancellation, Speech Enhancement and Speech Separation}
\name{Tom O'Malley, Arun Narayanan, Quan Wang, Alex Park, James Walker, Nathan Howard}
\address{Google LLC, U.S.A}
\begin{document}
\ninept
\maketitle
\begin{abstract}
We present a frontend for improving robustness of automatic speech recognition (ASR), that jointly implements three modules within a single model: acoustic echo cancellation, speech enhancement, and speech separation.
This is achieved by using a contextual enhancement neural network that can optionally make use of different types of side inputs: (1) a reference signal of the playback audio, which is necessary for echo cancellation; (2) a noise context, which is useful for speech enhancement; and (3) an embedding vector representing the voice characteristic of the target speaker of interest, which is not only critical in speech separation, but also helpful for echo cancellation and speech enhancement.
We present detailed evaluations to show that the joint model performs almost as well as the task-specific models, and significantly reduces word error rate in noisy conditions even when using a large-scale state-of-the-art ASR model. Compared to the noisy baseline, the joint model reduces the word error rate in low signal-to-noise ratio conditions by at least 71\% on our echo cancellation dataset, 10\% on our noisy dataset, and 26\% on our multi-speaker dataset. Compared to task-specific models, the joint model performs within 10\% on our echo cancellation dataset, 2\% on the noisy dataset, and 3\% on the multi-speaker dataset.
\end{abstract}
\begin{keywords}
Noise robust ASR, Speaker embedding, Neural AEC, VoiceFilter
\end{keywords}
\section{Introduction}
\label{sec:intro}

Robustness of automatic speech recognition (ASR) systems has significantly improved over the years with the advent of neural networks based end-to-end models~\cite{PrabhavalkarRaoSainathLiEtAl17,BattenbergChenChildCoatesEtAl17,HoriWatanabeZhangChan2017,li2021betterfaster}, large-scale training data~\cite{mirsamadi2017multi,hakkani2016multi,NarayananMisraSimPundakEtAl18}, and better data augmentation strategies~\cite{kim2017mtr, park2019specaugment, medennikov2018investigation}. Nevertheless, various factors like device echo (in the case of smart speakers), harsher background noise and competing speech, still significantly deteriorate performance~\cite{barker2017thirdchime,barker2018fifthchime}. While it is possible to train separate ASR models that specifically address these conditions, for practical reasons, it is harder to maintain multiple task-specific ASR models and switch on-the-fly based on the use case. Furthermore, with large scale multi-domain \cite{NarayananMisraSimPundakEtAl18} and multi-lingual modeling \cite{adams2019massively} gaining more research interest, the training data for ASR models often covers varied use cases (acoustic and linguistic), like voice search and video captioning, making it more challenging to simultaneously address harsher noise conditions. As a result, it is often convenient to train and maintain separate frontend feature processing models that handle adverse conditions, without combining it with the backend ASR model.

Background interference types can be broadly classified into 3 groups: Device echo, background noise, and competing speech. 

\paragraph{Device echo} For devices like smart home speakers that play audio, the device's echo, i.e., recording of the audio that is being played back, can affect performance. For an ASR system, degradation can be especially severe if audio being played back contains audible speech, which is often the case. Acoustic echo cancellation (AEC) techniques~\cite{hansler2005acoustic,benesty2001advances,Hu2006b} are used to address this. Both signal processing based~\cite{hansler2005acoustic, benesty2001advances, benesty2011perspective, enzner2014acoustic} and neural-net based~\cite{zhang2018deep, fazel2019deep, lei2019deep, ding2020textual} approaches have been proposed. A unique characteristic of AEC is that the reference signal being played back is typically available and can be used for suppression. The challenge arises in the fact that the signal-to-noise ratio (SNR) is usually quite low because the microphones are often close to the speakers, and the reference signal goes through device non-linearities and room acoustics, which make it challenging to directly subtract it from the signal recorded by the device. 

\paragraph{Background noise} Background noise with non-speech characteristics is usually well handled using data augmentation strategies like multi-style training (MTR) of the ASR models~\cite{lippmann1987multi,kim2017mtr}: A room simulator is used to add noise to the training data, which is then carefully weighted with clean data during training to get a good balance in performance between clean and noisy conditions. As a result, large scale ASR models are robust to moderate levels of non-speech noise. But background noise can still affect performance in low SNR conditions~\cite{ko2017study}.

\paragraph{Competing speech} Unlike non-speech noise, competing speech is quite challenging for ASR models that are trained to recognize a single speaker. Training ASR models with multi-talker speech can pose problems in itself, since it is hard to disambiguate which speaker to focus on during inference. Using models that recognize multiple speakers~\cite{tripathi2020end} is also sub-optimal since it is hard to know ahead of time how many users to support. Furthermore, such multi-speaker models typically have degraded performance in single-speaker settings, which is undesirable.

\begin{figure*}[tbh!]
	\centering
	\includegraphics[width=0.6\textwidth]{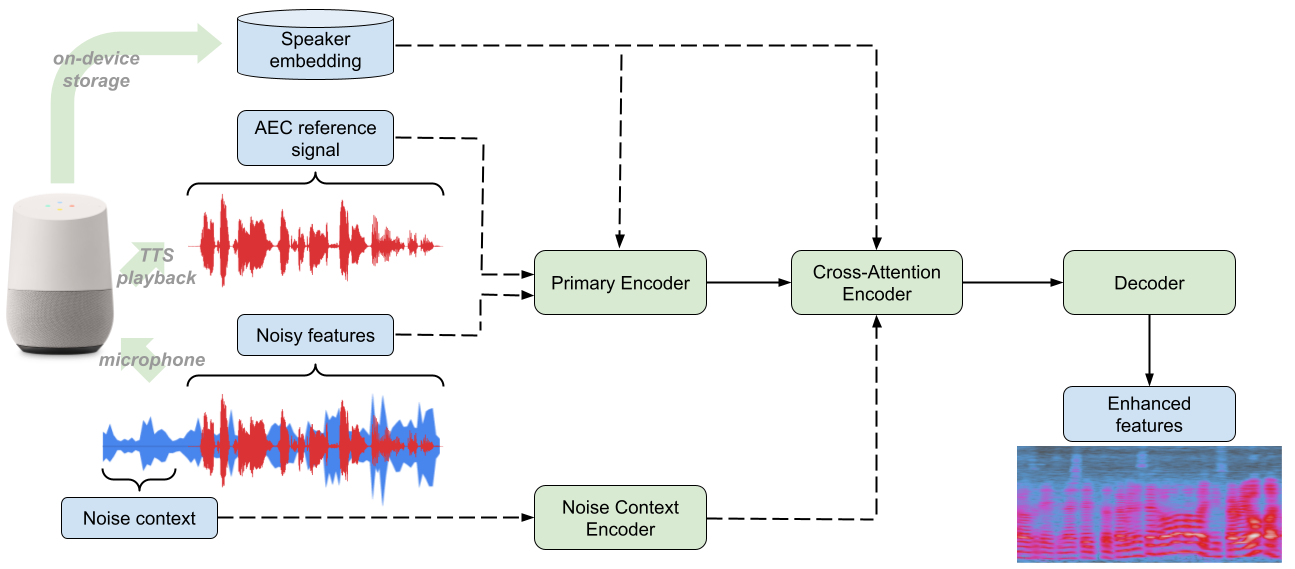}
	\caption{System overview. The system receives noisy features, and 3 optional signals: the AEC reference signal, the noise context, and the speaker embedding. The noisy features and AEC reference signal are concatenated and encoded together. The speaker embedding is used to modulate the encoded features of the system. The noise context is encoded separately, and then combined with the encoded features via cross-attention.}
	\label{fig:system_overview}
\end{figure*}

The three classes of interference mentioned above have been addressed in the literature, typically in isolation, using separate modeling strategies. Speech separation has received a lot of attention in the recent literature using techniques like deep clustering~\cite{hershey2016deep}, permutation invariant training~\cite{yu2017permutation,kolbaek2017multitalker}, and using speaker embedding~\cite{zmolikova2017speaker,wang2018deep,Wang2019}. When using speaker embedding, the target speaker of interest is assumed to be known \textit{a priori}. Techniques developed for speaker separation have also been applied to remove non-speech noise~\cite{maciejewski2020whamr,zeghidour2020wavesplit}, with modifications to the training data. AEC has also been studied in isolation~\cite{valin2020perceptually}, or together with background noise\cite{zhang2018deepAEC}. While most approaches focus on improving quality of enhanced speech, some have also focused on improving ASR~\cite{zhang2018deepAEC,howard2021neural}, as we do in this work. It is well known that improving speech quality does not always improve ASR performance~\cite{Seltzer2013DNNAurora4} since the distortions introduced by non-linear processing can adversely affect ASR. One way to mitigate this is to jointly train the enhancement frontend together with the backend ASR model~\cite{Narayanan2014Joint}.

In the presented work, we address all three interference types jointly. A joint model is interesting for practical reasons since it is hard to know the interference type to address ahead of time especially in a streaming recognition setting. The joint model is constructed as a \emph{contextual} frontend processing model, wherein the contextual information is assumed to be optional. In the case of AEC, a reference signal and a speaker embedding of the target speaker is assumed to be available. In the case of speech enhancement and separation, a noise context, i.e. a few seconds of audio before the target utterance to be recognized, and the target speaker embedding are assumed to be available. Noise context carries useful information about the acoustic context and has been shown to be useful in prior work \cite{huang2019hotwordcleaner}. A single model then processes these contextual signals to produce enhanced features that are passed to the ASR system. Reference and noise context, when not available, is assumed to be an uninformative silence signal.

The backbone of the contextual frontend model is the recently proposed conformer architecture~\cite{gulati2020conformer}, which has been shown to be especially well-suited for speech tasks like ASR~\cite{gulati2020conformer} and enhancement~\cite{chen2021continuous}. We extend this architecture to make use of reference signal, speaker embedding, and noise context simultaneously. Our results show that a joint model can work almost as well as task specific models; AEC performance is within 10\% of a standalone AEC model, and 71\% better than the baseline. Performance in the presence of background noise is 10\% better than the baseline and 2\% of a standalone enhancement model, while in the presence of competing speech, it is 26\% better than the baseline and within 3\% of a standalone system.

The rest of the paper is organized as follows. {Section~\ref{sec:method}} presents various components of the joint model in detail. {Section~\ref{sec:experiments}} and {Section~\ref{sec:results}} present our experimental settings and detailed results and analysis. Concluding remarks and a discussion of future directions are presented in {Section~\ref{sec:results}}.

\section{System Description}
\label{sec:method}

\subsection{Contextual inputs}
\subsubsection{Acoustic Echo Cancellation}

In order to perform acoustic echo cancellation, we use the reference signal that is being played back by the device as an input to the system. It is assumed that the reference signal is temporally aligned with the input, and is of the same length. We compute log Mel-filterbank energy (LFBE) features of the reference signal, and stack it with LFBE features of the input signal. The stacked features are used as input to the enhancement frontend. When there is no signal being played by the device, we use an all-zero reference signal.

\subsubsection{Noise Context Modeling}

As noise context, the system makes use of a 6 seconds noise segment before target speech. When generating training and evaluation sets, we select noise segments longer than the target utterance, and segment out the first 6 seconds to be used as noise context. LFBE features of the noise context signal are used as context information. They are used by the model using a specialized architecture explained in greater detail in Sections~\ref{sec:noise-context-encoder} and \ref{sec:cross-attention-encoder}.

\subsubsection{Target Speaker Modeling}

The speaker embeddings (\emph{a.k.a.} d-vectors) are computed using a text-independent speaker recognition model trained with the generalized end-to-end extended-set softmax loss~\cite{wan2018generalized,pelecanos2021dr}.
The model has 3 LSTM layers each with 768 nodes and a projection size of 256. The output of the final frame of the last LSTM layer is then linearly transformed to the final 256-dimension d-vector.

For training and evaluations, each target utterance is paired with a separate ``enrollment'' utterance from the same speaker. The enrollment utterance is selected at random from a pool of available utterances of the target speaker. The d-vectors are then computed on the enrollment utterance.
For most real applications, the enrollment utterances are usually obtained via a separate offline process~\cite{enrollmentblog,wang2020version}.

\subsection{Model architecture}
\label{sec:subsubhead}

The proposed joint frontend model uses a modified version of the conformer architecture~\cite{gulati2020conformer}, which combines convolution and self-attention to model short-range and long-range interactions, respectively. This architecture has been shown to be well-suited for modeling speech and audio in prior work \cite{gulati2020conformer, chen2021continuous}. A conformer layer consists of a sequence of half feed-forward net, a convolutional block, a self-attention block, and a final half feed-forward net, along with residual connections and layer norms in between these blocks. We modify the conformer architecture to allow using d-vector and noise-context as additional side inputs.

Fig.~\ref{fig:system_overview} shows an overview of the overall architecture. The system is composed of a primary encoder, a noise context encoder, and a cross-attention encoder. In all cases, when a conformer block is used, we use local, causal self-attention, to allow the model to be used for streaming. As shown in the figure, the primary encoder gets the stacked LFBE features from the noisy signal and the AEC reference, along with the d-vector. The noise-context encoder, receives LFBE features from the noise context as its input. The output of the primary encoder and the noise-context encoder are passed as the main input and the auxiliary input, respectively, to the cross-attention encoder, along with the d-vector of the target speaker. Finally. the output of the cross-attention encoder is passed through a simple feed-forward network to predict the enhancement targets. Each of these modules are described in the following subsections.

\subsubsection{Primary Encoder}
\label{sec:primary-encoder}

\begin{figure}[bth]
	\centering
	\includegraphics[width=0.45\textwidth]{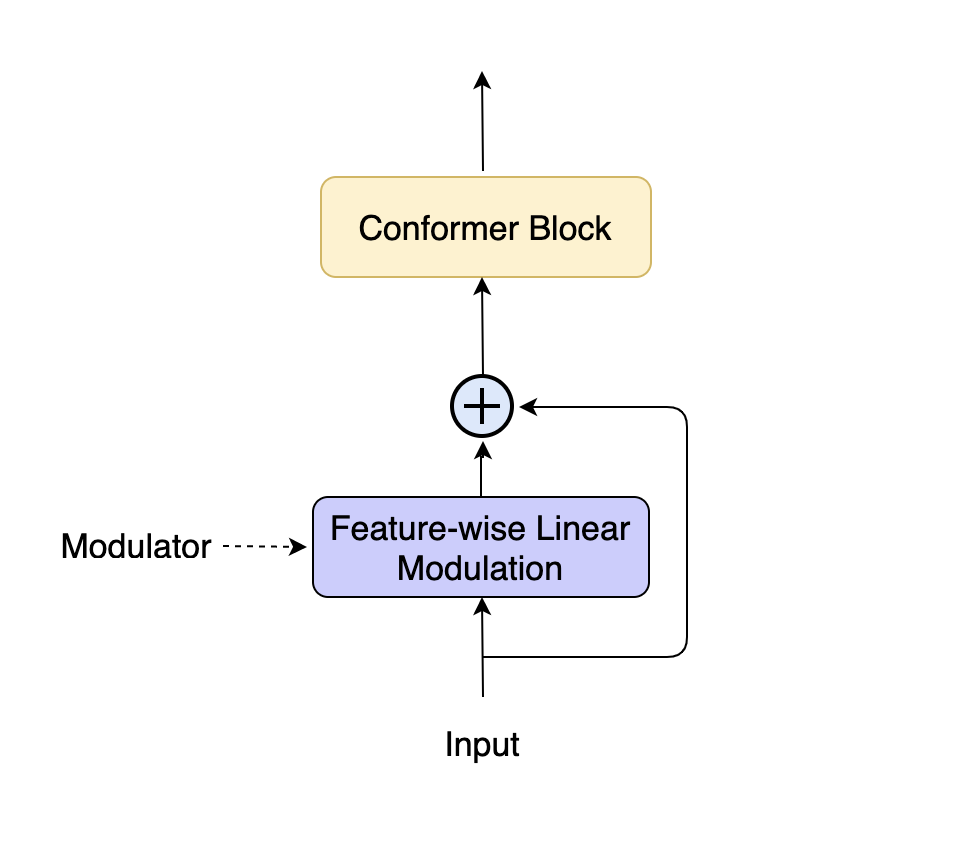}
	\caption{Modulated Conformer Block. The input features are modulated with an optional auxiliary signal, in our case the speaker embedding vector. The modulation occurs via a FiLM layer~\cite{perez2018film} with a residual connection. The modulated input features are then passed to the Conformer block.}
	\label{fig:modulated_conformer}
\end{figure}

The primary encoder consists of $N$ modified conformer blocks. The inputs to the primary encoder are LFBE features from the noisy signal and the AEC reference signal. The features are stacked together before being passed to the encoder. The encoder also recieves the d-vector of the target speaker as additional input. Before each conformer block, the d-vector is combined with the block's inputs using feature-wise linear modulation (FiLM) \cite{perez2018film}. This allows the primary encoder to adjust its encoding based on the embedding of the target speaker. A residual connection is added after the feature-wise linear modulation layer, in order to ensure that the architecture can perform well when the speaker embedding is absent. The rest of each block follows the pattern of a standard conformer block. We refer to the resulting architecture as a \textit{modulated conformer block}. A representation of this block is shown in Fig. ~\ref{fig:modulated_conformer}. Mathematically, the modulated conformer block transforms its input features $x$, using modulation features $m$, to produce output features, $y$, as follows:

\begin{align}
\tilde{x} &= x + r(m) \odot x + h(m), \label{eq:film}\\
x^{\prime} &= \tilde{x} + \frac{1}{2}\textrm{FFN}(\tilde{x}), \nonumber\\
x^{\prime\prime} &= x^{\prime} + \textrm{Conv}(x^{\prime}),  \nonumber\\
x^{\prime\prime\prime} &= x^{\prime\prime} + \textrm{MHSA}(x^{\prime\prime}),  \nonumber\\
y &= \textrm{LayerNorm}(x^{\prime\prime\prime} + \frac{1}{2} \textrm{FFN}(x^{\prime\prime\prime})).  \nonumber
\end{align}
Here, $h(\cdot)$ and $r(\cdot)$ are affine transformations. FFN, Conv, and MHSA stand for feed-forward module, convolution module, and multi-headed self-attention module, respectively. Eq.~\ref{eq:film} shows the feature-wise linear modulation (FiLM) layer, with a residual connection. The rest follows the formulation in a standard conformer block.

\subsubsection{Noise Context Encoder}
\label{sec:noise-context-encoder}

The noise context encoder also consists of N conformer blocks, and can be executed in parallel with the primary encoder. The inputs to the noise context encoder are the features of the noise context, utilizing the noise present up to 6 seconds prior to the target utterance. The encoder uses LFBE features from the noise context as input. It employs standard conformer blocks, without modulation by the speaker embedding. This is because the noise context prior to the target utterance is assumed to contain information that we want to pass forward to the cross-attention encoder (Sec.~\ref{sec:cross-attention-encoder}), to aid noise suppression.

\subsubsection{Cross-Attention Encoder}
\label{sec:cross-attention-encoder}

The outputs of the primary encoder are fed to the cross-attention encoder \cite{narayanan2021crossattention}. The cross-attention encoder is composed of $M$ modified cross-attention conformer blocks, which we refer to as \textit{modulated cross-attention conformer} blocks. This block receives the output of the primary encoder as its main input, and the output of the noise-context encoder as the auxiliary input. The main input is first modulated using the d-vector, using feature-wise linear modulation as described in Sec.~\ref{sec:primary-encoder}. The cross-attention conformer first independently processes the modulated input and the auxiliary input using half feed-forward nets, and convolutional blocks. Subsequently, a cross-attention block is used to summarize the auxiliary inputs, using the processed input as the query vectors. Intuitively, the role of the cross-attention block is to summarize noise context separately for each input frame that is to be enhanced. The summarized auxiliary features are then merged with the inputs using a FiLM layer, which is followed by a second cross-attention layer to further process the merged features. Mathematically, if $x$, $m$, and $n$ are the encoded input, d-vector and the encoded noise context from the previous layer, cross attention encoder does the following:
\begin{align}
\begin{split}
\hat{x} &= x + r(m) \odot x + h(m),\\
\tilde{x} &= \hat{x} + \frac{1}{2}\textrm{FFN}(\hat{x}), \tilde{n} = n + \frac{1}{2}\textrm{FFN}(n),  \\
x^{\prime} &= \tilde{x} + \textrm{Conv}(\tilde{x}), n^{\prime} = \tilde{n} + \textrm{Conv}(\tilde{n}), \\
x^{\prime\prime} &= x^{\prime} + \textrm{MHCA}(x^{\prime}, n^{\prime}), \\
x^{\prime\prime\prime} &= x^{\prime} \odot r(x^{\prime\prime}) + h(x^{\prime\prime}), \\
x^{\prime\prime\prime\prime} &= x^{\prime} + \textrm{MHCA}(x^{\prime}, x^{\prime\prime\prime}), \\
y &= \textrm{LayerNorm}(x^{\prime\prime\prime\prime} + \frac{1}{2} \textrm{FFN}(x^{\prime\prime\prime\prime})). \\
\end{split}
\end{align}
Thus the inputs are modulated at each block by both the target speaker's embedding and the encoded noise context.

\subsubsection{Decoder}

We use a simple projection decoder. The projection decoder consists of a single-layer, frame-wise fully connected network with sigmoid activation.

\subsection{Training target}

We use the ideal ratio mask (IRM) as the training target \cite{Narayanan2013IRM}. IRMs are computed using reverberant speech and reverberant noise. This computation assumes that speech and noise are uncorrelated in Mel spectral space:
\begin{equation}
\label{eq_irm_def}
\IRM(\TI,\FI) = \frac{\cleanCgm(\TI,\FI)}{\cleanCgm(\TI,\FI) + \noiseCgm(\TI,\FI)}.
\end{equation}
Here, $\cleanCgm$ and $\noiseCgm$ are the reverberant speech and reverberant noise Mel-filterbank energies, respectively. $\TI$ and $\FI$ represent time and Mel frequency bin indices. Using the IRM allows us to do enhancement directly in the feature space for ASR, without any need for reconstructing the waveform. This also allows the enhancement frontend to be trained together with the ASR model, which is crucial for improving performance as we'll show in Sec.~\ref{sec:results}.

\subsection{Loss}

We combine two losses during training: a spectral loss and a loss based on the outputs of an ASR encoder.

\subsubsection{Spectral Loss}

The spectral loss consists of the $\ell$1 and $\ell$2 distance between the estimated ratio mask and the ideal ratio mask.

\subsubsection{ASR Loss}

The ASR loss is computed by passing LFBE features of the target utterance and those computed by the proposed enhancement frontend to a pre-trained end-to-end ASR model. Following \cite{howard2021neural}, we only use the encoder of the ASR model for computing the loss. The loss is computed as the $\ell$2 distance between the outputs of the ASR encoder for the target features and the enhanced features. The ASR encoder is not updated during training. The goal of using ASR loss is to make enhancement be more attuned to the ASR model, which is critical for getting the best performance out of the enhancement frontend. By keeping the ASR model's parameters fixed, we also decouple the ASR model and the enhancement frontend, thereby allowing them to be trained and deployed independent of each other.

\subsection{Inference}

During inference, we scale the estimated ratio masks, before using it to enhance the LFBE features. This limits the amount of speech distortion introduced by masking, at the expense of reduced noise suppression. Prior work has shown that scaling the mask estimate improves ASR performance, since the ASR model is sensitive to distortions introduced by masking process \cite{Narayanan2013IRM}. The mask is applied before log compression of the features, as:
\begin{equation}
\label{eq:masking}
\EstCleanCgm(\TI,\FI) = \mixCgm(\TI,\FI) \odot \max(\ERM(\TI,\FI), \beta)^\alpha.
\end{equation}
Here, $\alpha$ and $\beta$ are the mask scalar and mask floor, $\ERM$ is the estimated mask, $\mixCgm$ is the noisy Mel-filterbank features, $\EstCleanCgm$ is an estimate of clean Mel-filterbank features, and $\odot$ is pointwise multiplication. In all our experiments, $\alpha$ and $\beta$ are set to 0.5 and 0.01, respectively.

\section{Experimental settings}
\label{sec:experiments}

\begin{table*}[th]
  \centering
  \caption{WERs for baseline and AEC-only models.}
  \label{table:aec-results}
  \vspace{0.05in}
  \begin{tabular}{c|cc|cccc}
   \multirow{2}{*}{Models}  & \multirow{2}{*}{Modulation} & ASR &
   \multicolumn{4}{c}{Librispeech-AEC} \\
   & & Loss & -10 dB & -5 dB & 0 dB & 5 dB \\ \hline
    Baseline      & - & - & 80.5 & 72.7 & 58.0 & 36.1 \\ \hline
    AEC-only     & None & N & 51.1 & 30.3 & 17.8 & 12.2 \\ %
    AEC-only    & Linear & N & 51.9 & 30.9 & 18.0 & 12.4 \\ %
    AEC-only     & FiLM & N & 51.3 & 30.8 & 17.9 & 12.3 \\ %
    AEC-only     & None & Y & 21.4 & 14.7 & 11.3 & \textbf{9.4} \\ %
    AEC-only     & Linear & Y & 21.2 & 14.6 & \textbf{11.2} & \textbf{9.4} \\ %
    AEC-only     & FiLM & Y & \textbf{21.0} & \textbf{14.5} & 11.3 & \textbf{9.4} \\ \hline
  \end{tabular}
  \vspace{-0.05in}
\end{table*}

\begin{table*}[th]
  \centering
  \caption{WERs for baseline and Enhancement-only models.}
  \vspace{0.05in}
  \begin{tabular}{c|cc|ccc|ccc}
   \multirow{2}{*}{Models}  & \multirow{2}{*}{Modulation} & ASR & 
   \multicolumn{3}{c}{Librispeech-Multispeaker} & \multicolumn{3}{c}{Librispeech-Noise} \\
   & & Loss & -5 dB & 0 dB & 5 dB & -5 dB & 0 dB & 5 dB \\ \hline
Baseline & - & - & 69.2 & 46.4 & 29.3 & 36.5 & 22.5 & 14.0 \\ \hline
Enh-only & None & N & 51.3 & 33.1 & 22.3 & 33.3 & 20.4 & 13.1 \\
Enh-only & Linear & N & 46.8 & 31.6 & 21.8 & 32.6 & 19.8 & 13.0 \\
Enh-only & FiLM & N & 46.8 & 31.5 & 21.7 & 32.5 & 19.9 & 13.1 \\
Enh-only & None & Y & 52.0 & 33.1 & 22.3 & 32.8 & 20.1 & 13.1 \\
Enh-only & Linear & Y & 46.7 & 31.5 & 21.9 & 32.1 & 19.8 & 13.0 \\
Enh-only & FiLM & Y & \bf 45.3 & \bf 30.4 & \bf 21.4 & \bf 30.6 & \bf 18.9 & \bf 12.5 \\
  \end{tabular}
  \label{table:enhancement-results}
  \vspace{-0.05in}
\end{table*}

\begin{table*}[th]
  \centering
  \caption{WERs for baseline, joint, and single-task models.}
  \vspace{0.05in}
  \begin{tabular}{c|cc|ccc|ccc|cccc|cc}
   \multirow{3}{*}{Models}  & \multirow{3}{*}{Modulation} & ASR & \multicolumn{12}{c}{Librispeech} \\
   & & Loss & \multicolumn{3}{c}{Multi-speaker} & \multicolumn{3}{c}{Noisy}  &
   \multicolumn{4}{c}{AEC} & Test & Test \\
   & & & -5 dB & 0 dB & 5 dB & -5 dB & 0 dB & 5 dB  & -10 dB & -5 dB & 0 dB & 5 dB & Clean & Other \\ \hline
Baseline & - & - & 69.2 & 46.4 & 29.3 & 36.5 & 22.5 & 14.0 & 80.5 & 72.7 & 58.0 & 36.1 & \bf 6.7 & \bf 12.8 \\ \hline
Joint & None & Y & 49.8 & 31.8 & 21.5 & 31.3 & 19.3 & 12.7 & 22.8 & 15.7 & 11.7 & 9.7 & 6.8 & 12.9 \\
Joint & Linear & Y & 46.9 & 31.5 & 22.1 & 31.6 & 19.3 & 12.9 & 23.0 & 15.8 & 12.0 & 9.8 & \bf 6.7 & \bf 12.8 \\
Joint & FiLM & Y & 46.6 & 31.3 & 21.8 & 31.3 & 19.2 & 12.6 & 23.1 & 15.6 & 11.9 & 9.8 & \bf 6.7 & \bf 12.8 \\ \hline
Enh-only & FiLM & Y & \bf 45.3 & \bf 30.4 & \bf 21.4 & \bf 30.6 & \bf 18.9 & \bf 12.5 & - & - & - & - & \bf 6.7 & \bf 12.8 \\ \hline
AEC-only & FiLM & Y & - & - & - & - & - & - & \bf 21.0 & \bf 14.5 & \bf 11.3 & \bf 9.4 & \bf 6.7 & \bf 12.8 \\
  \end{tabular}
  \label{table:joint-results}
  \vspace{-0.05in}
\end{table*}

We train 3 sets of models: separate speech enhancement and AEC models, and a joint model that performs both. All three models incorporate the dvector in order to perform speaker separation. The speech enhancement model is trained solely on our speech enhancement dataset, the AEC model is trained solely on our AEC dataset, and the joint model is trained on a mixture of the two datasets. We also perform an ablation study to measure the effects of removing the ASR loss and the speaker embedding from the models.

\subsection{Datasets}

The datasets for speech enhancement and AEC are derived from LibriSpeech~\cite{librispeech} and internal vendor-collected utterances. LibriSpeech consists of $281$k utterances, and the vendor collected sets consist of $1,916$k utterances. To generate noisy data, we consider these utterances as `clean', and add noise to it.

\subsubsection{Speech Enhancement}
The speech enhancement training set is created by passing the clean utterances through a room simulator \cite{kim2017mtr}, that first adds reverberation, followed by noise. The signal to noise ratio (SNR) is in the range of [-10~dB, 30~dB]. The reverberation time is set to be between 0 msec and 900 msec. The noise snippets represent typical noise conditions like cafeteria, kitchen, cars, etc. We also include publicly available noises from Getty Audio\footnote{\url{https://www.gettyimages.com/about-music}} and YouTube Audio Library\footnote{\url{https://youtube.com/audiolibrary}}. In order to simulate multi-talker conditions, we also mix the the training utterances with competing speech from the training datasets, chosen randomly. To increase the diversity of the training data, we generate multiple copies of each utterances under different noise and room conditions. 

For evaluating speech enhancement performance, we create 2 sets of test sets, one with background noise and one with competing speech. The target speech comes from the Librispeech test sets. The background noise test sets are created similar to the training set, but using held-out noise snippets, and at SNRs -5~dB, 0~dB and 5~dB. The multi-speaker test sets are also created similar to the corresponding training set, at SNRs -5~dB, 0~dB and 5~dB. 

\subsubsection{Acoustic Echo Cancellation}

Motivated by findings in \cite{howard2021neural}, we created two types of AEC training data.  The first type consists of entirely synthetic echoes.  A reference signal is played out through a reverberant room simulator with the speaker configured close to the microphone(s).  We mix this synthetic echo with a reverberant target speech signal at SNRs ranging from $-20~$dB to $5$~dB.  Both target speech and reference signals are drawn from the Librispeech training set  when generating this type of data. The second type of training data consists of re-recorded echoes.  For this subset, we re-record utterances drawing from a dataset collected internally for text-to-speech (TTS) purposes.  We additionally augment this set with actual TTS utterances. These utterances are played back in a room on Google Home devices at varying signal levels and re-recorded. The goal is to capture microphone non-linearities, which are typically harder to model using a room simulator. The re-recorded reference signals are then added to target utterances, which are themselves reverberated using a room simulator, at SNRs ranging from $-20~$dB to $5$~dB.

The AEC evaluation sets are created using only re-recorded echoes, since that subset is both more challenging and closer to real-world AEC conditions.  The target speech comes from Librispeech test-clean set, to which, after adding reverberation, we add re-recorded echo at SNRs from -10~dB to 5~dB in 5~dB increments. 

\vspace{-0.1in}
\subsubsection{Librispeech Test and Other}

We also test our best single-task and joint models on the Librispeech test-clean and test-other datasets without any additional noise added. We do this to ensure that our frontend does not deteriorate the performance of the downstream ASR model in high SNR conditions.

\subsection{Training Details}

The models are trained in TensorFlow \cite{abadi2016tensorflow} with the Adam optimizer \cite{kingma2014adam}. We use the open-source Lingvo \cite{shen2019lingvo} Conformer encoder. LFBE features with a dimension of 128 are used for the target utterance, noisy utterance, noise context signal, and reference signal. The dvector is 256-dimensional. Conformer blocks use causal, masked self-attention, with a window of 65 frames in the past. The conformer blocks input and output 256-dimensional features. The hidden layers of the feed-forward module of the conformer blocks use a dimensional multiplier of 6 for the enhancement-only and joint models, and a multiplier of 8 for the AEC-only model. All models contain approximately 15M parameters.

The AEC-only model does not incorporate noise context. As such, it uses 6 modulated conformer blocks in the primary encoder, and does not contain a noise context encoder or cross-attention encoder. The enhancement-only model and the joint model both use 2 modulated conformer blocks for the primary encoder, 2 conformer blocks for the noise context encoder, and 2 modulated cross-attention conformer blocks for the cross-attention encoder.

When using ASR loss, the enhancement model is first trained for 20k steps with only the spectral loss. We then increase the weighting of the ASR loss linearly over the next 80k steps for the AEC-only and the enhancement-only models. For the joint model, ASR loss weight is increased more gradually over the next 180k steps. We found the slower ramp-up to improve training stability.

The ASR model that is used for evaluation is a recurrent neural net transducer model  with LSTM based encoder layers \cite{sainath2020streaming}. The training data for this model comes from varied domains that include VoiceSearch, Farfield, Telephony and YouTube. The utterances are anonymized and hand transcribed, and totals to $\sim$400k hours of speech \cite{narayanan2019longform}. Additionally, we use a room simulator to generate noisy versions of these datasets at SNRs ranging from 0~dB to 30~dB, and reverberation times ranging from 0 msec to 900 msec. The model is quite robust to moderate levels of noise, as can be seen from the results in Sec.~\ref{sec:results}. The features used by the ASR model matches those used by the enhancement frontend: 128-dimensional LFBE  features computed for 32 msec windows with 10 msec hop. But unlike the frontend, the ASR model stacks 4 contiguous frames of features and subsamples them over time by a factor of 3. Therefore, when using the frontend to enhance features, the enhanced LFBE are similarly stacked and subsampled before being passed to the ASR model. Note that even though the enhancement frontends are jointly trained with the ASR model, the ASR model parameters remain unchanged. 

We use word error rate (WER) as the evaluation metric. The WER is computed using the model described above, with and without the enhancement frontends.

\subsection{Ablation study}

In addition to comparing the performance of our joint model with models trained for only AEC or enhancement, we select two characteristics of our model for an ablation study. First, we compare the performance of our model with and without ASR loss. Second, we compare the performance of our model with and without incorporating dvector. We also compare the results of using a linear projection layer, rather than feature-wise linear modulation, when incorporating the speaker embedding into our encoders.

\section{Results}
\label{sec:results}

In the following section, we examine the results of our experiments, and compare the WER we achieve with a baseline in which the noisy LFBE features are fed directly to an ASR model. 

\subsection{AEC-only Results}

Table \ref{table:aec-results} shows the results of our experiments on the Librispeech AEC dataset. ASR loss showed a dramatic improvement in the performance of the models, with the best AEC-only model without ASR loss yielding a WER of 51.1\% at an SNR of -10~dB, compared to 21.0\% when trained with ASR loss. This represents a 58\% reduction in the relative WER. At low SNRs, the IRM has fewer time-frequency units that are dominated by target speech, and therefore has more values closer to 0.0. ASR loss likely helps retain the most critical information in the estimated IRM that are relevant for ASR, thereby improving WER by such a large margin. As the input SNR increases, the relative gains by adding ASR loss also goes down. At 5~dB, ASR loss improves performance by 23\% relative.  

Modulating with d-vector marginally improved performance, reducing WER by 2\% at -10 dB. This is likely because the reference signal provides enough information to separate it out from the target signal, making dvector information partially redundant.

\subsection{Enhancement-only Results}

Table \ref{table:enhancement-results} shows the results of our enhancement experiments on the Librispeech multi-speaker and Librispeech noisy datasets. For these tasks, both d-vector modulation and ASR loss improves performance. As expected, the improvement due to d-vector was most noticeable on the multi-speaker datasets. For this datasets, the best model with d-vector reduced the WER at SNR -5 dB by 11.7\% compared to the best model without d-vector. 

For both datasets, at all SNR's, the model incorporating the d-vector via feature-wise linear modulation and trained with ASR loss performed the best. In all cases, the incorporation of the d-vector out-performed or equaled the performance of the comparable model without the d-vector. The average relative WER improvement at SNR -5 dB was 7.6\%. The addition of ASR loss provided more gains when paired with modulation by the d-vector. When paired with FiLM, ASR loss improved performance by 3.2\% and 6.1\% on the multi-speaker and noisy datasets, respectively. In other cases, the ASR loss improved performance only marginally.

\subsection{Joint Results}

Table \ref{table:joint-results} compares the performance of our joint model on these tasks with the performance of our best single-task models. For the Librispeech multi-speaker dataset, the joint model performs within 3\% of the best single-task model, and reduces WER by 33\% compared to baseline at a SNR of -5 dB. For the Librispeech noisy dataset, the joint model performs within 2\% of the best single-task model, and reduces WER by 14\% compared to baseline at -5 dB SNR. For the Librispeech AEC dataset, the joint model performs within 10\% of the best single-task model, and reduces WER by 71\% compared to baseline at the most challenging SNR of -10 dB. The small drop in performance is expected, because the joint model must learn 3 tasks with roughly the same number of parameters as each single-task model. Additionally, the results on Librispeech test and Librispeech other show that the performance of the ASR model on clean data did not degrade when this data is first passed through our frontend. This ensures we do not sacrifice performance in favorable environments by always running the frontend.

\section{Conclusion}

In this work, we presented a conformer-based architecture for joint AEC, speech enhancement, and speech separation. We incorporate three optional contextual features into this model: a reference signal from the device playback, noise context, and a speaker embedding. Since the reference signal is the same length as the utterance, it is directly concatenated with the input features before these features are fed to the encoder. Meanwhile, we use cross-attention to incorporate the noise context, since it is a sequential feature that may differ in length from the utterance. Finally, we use feature-wise linear modulation to incorporate the speaker embedding, since this is a global context feature.

We showed how the joint model can improve the performance of a downstream ASR model. Compared to passing the noisy features directly to an ASR model, our joint model reduces WER in low SNR environments by 71\%, 14\%, and 33\% on AEC, speech enhancement, and speech separation datasets, respectively. In addition, we showed that a joint model can perform competitively with single-task models. Our joint model performed within 10\% of our best single-task model for AEC, 2\% on speech enhancement, and 3\% on speech separation.  

In future work, we will explore at a finer scale how each contextual signal interacts with the rest in the model, and how the model behaves when some of them are missing or corrupt. It would also be interesting to consider improving the model with other contextual signals, such as video.

\label{sec:conclusion}

\section{Acknowledgements} 
\label{sec:ack}

We thank Alex Gruenstein, Mert Saglam, and Sankaran Panchapagesan for several useful discussions, feedback, and help with datasets.

\clearpage
\bibliographystyle{IEEEbib}
\bibliography{refs}

\end{document}